\documentclass[aps,pra,amsmath,twocolumn,superscriptaddress,amssymb]{revtex4}

\usepackage{bbm}
\usepackage{graphicx}% Include figure files
\usepackage{dcolumn}% Align table columns on decimal point
\usepackage{bm}% bold math
\usepackage{amsmath}
\usepackage{amssymb,amsthm}
\usepackage{color}

\usepackage[sans]{dsfont}

\newcommand{\ket}[1]{|#1\rangle}
\newcommand{\braket}[2]{\langle{#1}|{#2}\rangle}

\newcommand{\bra}[1]{\langle#1|}

\def\eq{\begin{eqnarray}}
\def\en{\end{eqnarray}}

 \usepackage[normalem]{ulem}

\begin{document}

\title{Is macroscopic entanglement a typical trait of many-particle quantum states?} 

\author{Malte C. Tichy}
\address{Department of Physics and Astronomy, University of Aarhus, DK--8000 Aarhus C, Denmark}
\author{Chae-Yeun Park}
\address{Center for Macroscopic Quantum Control, Department of Physics and Astronomy, Seoul National University, Seoul 151-742, Korea}
\author{Minsu Kang}
\address{Center for Macroscopic Quantum Control, Department of Physics and Astronomy, Seoul National University, Seoul 151-742, Korea}
\author{Hyunseok Jeong}
\address{Center for Macroscopic Quantum Control, Department of Physics and Astronomy, Seoul National University, Seoul 151-742, Korea}
\author{Klaus M{\o}lmer}
\address{Department of Physics and Astronomy, University of Aarhus, DK--8000 Aarhus C, Denmark}

\begin{abstract}
We elucidate the  relationship between  Schr\"o{}dinger-cat-like macroscopicity and geometric entanglement and argue that these quantities are not interchangeable. 
While both properties are lost due to decoherence, we show that macroscopicity is rare in uniform and in so-called random  physical ensembles of pure quantum states, despite  possibly large geometric entanglement. 
 In contrast,  permutation-symmetric pure states feature rather low geometric entanglement and strong and robust macroscopicity. 
\end{abstract}
\date{\today}

\maketitle

\section{Introduction}
Quantum entanglement entails two important consequences:  On the one hand, it is the ``characteristic trait of quantum mechanics'' \cite{charactertraitSchr} that thoroughly thwarts our every day's intuition, most bizarrely when applied to macroscopic objects, as illustrated by the famous paradox of Schr\"odinger's cat \cite{Schrodinger1935}. 
On the other hand, entanglement is the very ingredient that makes the simulation of quantum many-body systems extremely challenging: A separable system of $N$ qubits requires only $2N$ parameters for its description, whereas an entangled state comes with $\sim 2^{N}$ variables. This curse of dimensionality in the context of simulating quantum systems becomes a powerful resource when it comes to the speedup of quantum computers over classical architectures. Although the two aspects of entanglement are two sides of the same medal, they are quantified differently: Measures of macroscopicity \cite{Frowis2012b} reflect the degree to which a quantum state resembles a Schr\"odinger's cat, the complexity of a quantum state is reproduced by the geometric measure of entanglement \cite{Shimony1995}, defined via the largest overlap to separable states. Macroscopicity can increase under local operations combined with classical communication (LOCC) \cite{Frowis2012b} in contrast to geometric entanglement, e.g.~the modestly macroscopic cluster state can be converted via LOCC into a highly macroscopic GHZ-state \cite{Briegel2001}.

In Nature,  macroscopic entanglement does not occur \cite{Leggett1980,Leggett2002} outside artificially tailored situations \cite{0953-4075-46-10-104001}. Its empirical absence contrasts with its immediate appearance in the formalism of quantum physics, which raises the question whether some unavoidable mechanism jeopardizes macroscopically entangled quantum states. The decoherence  programme \cite{Myatt2000} explains the emergence of classical behavior in our everyday's world and how macroscopic quantum superpositions decohere on overwhelmingly short timescales \cite{Zurek2003,Schlosshauer2005,Buchleitner:2009fk,Breuer:2006ud}:  The interaction between any quantum system and its surrounding environment destroys the coherence between macroscopically distinct alternatives -- such as the dead and the living cat. Even if the environment were shielded off perfectly, however, it remains unlikely that macroscopic entanglement be observed in Nature, due to the unavoidable coarse-graining of any measurement \cite{Raeisi2011,Sekatski2014,Sekatski2014a}. In other words, there are powerful mechanisms that quickly deteriorate macroscopic entanglement.

But the decoherence programme does not make any statement on the likelihood that a macroscopic quantum superposition may form, only that, \emph{when} it appears, it decoheres on a timescale so short that any attempt for observation is vain. Would a hypothetical decoherence-free world then host cohorts of Schr\"odinger's cats? In other words, how likely are macroscopically entangled quantum states \emph{before} the onset of decoherence? 

\begin{figure}[ht]
\includegraphics[width=.8\linewidth]{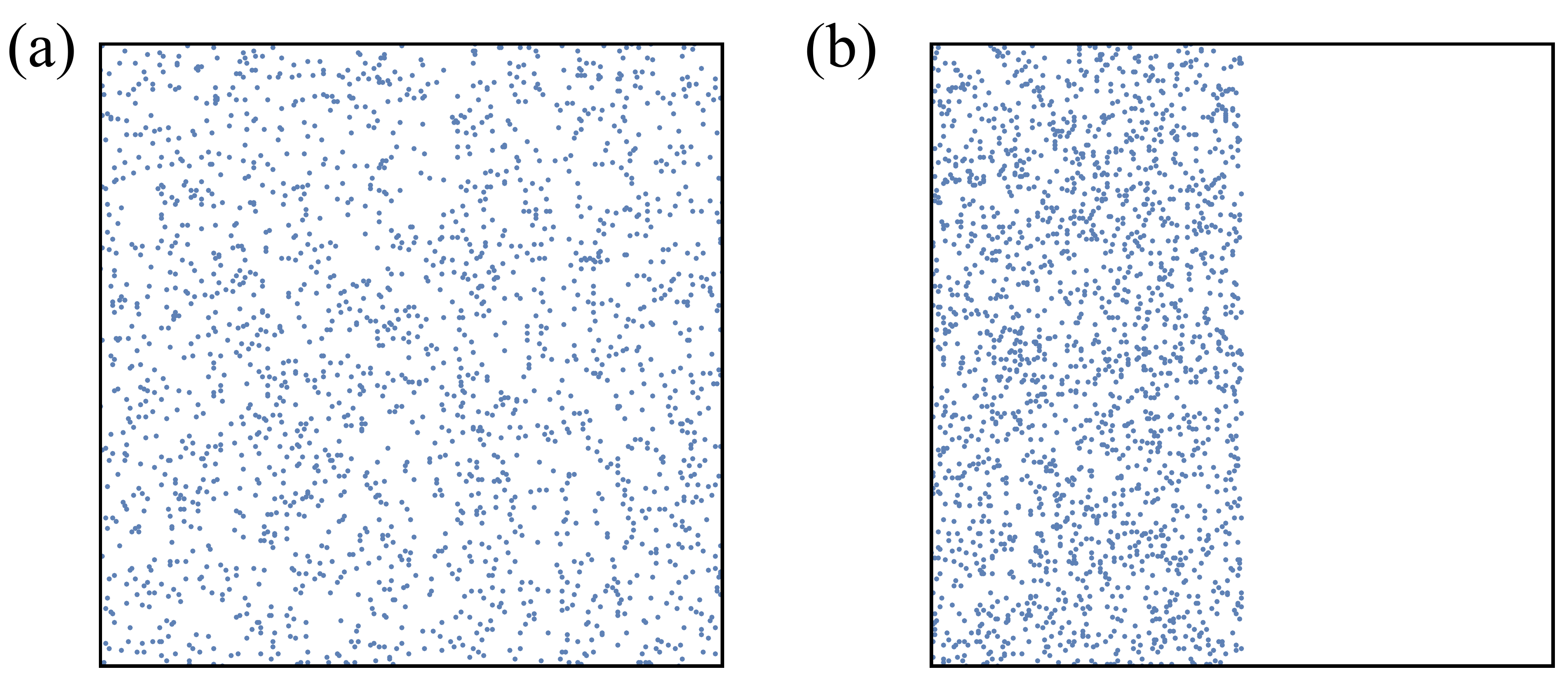}
\caption{Two microstates of the microcanonical ensemble of 2000 particles in a box. (a) Typical states feature a rather homogeneous distribution of particles, while (b) states with large inhomogeneities are artificial and rare. We argue in this article that the rarity of macroscopically entangled pure quantum states can be understood in close analogy: Macroscopically entangled states play the role of states with large inhomogeneities (b), and therefore form only extremely seldom. }
\label{analogy}
\end{figure}

In order to ease our intuition for these questions, let us propose a purely classical analogy: Consider the microcanonical ensemble of a gas of $N$ particles in a box  with total energy $E$, illustrated in Fig.~\ref{analogy}. In classical statistical mechanics, all microstates (specified by the positions and momenta of the gas particles $\{ \vec x_1, \dots , \vec x_N; \vec p_1, \dots , \vec p_N\}$) that are compatible with the total energy $E$ are assigned the  same probability \cite{huangstati}. How likely is it to find all particles in the left half of the box, as illustrated in Fig.~\ref{analogy}(b)? In the first place, if we prepared the particles in such a state, the system would relax to a homogenous distribution [Fig.~\ref{analogy}(a)] on a very short timescale -- just like decoherence  destroys any macroscopic quantum superposition. 

This explanation for the absence of the strongly inhomogeneous situation 
 is, however, not the commonly adopted one: The spontaneous occurrence of the macrostate sketched in Fig.~\ref{analogy}(b) is by itself extremely unlikely, and the relaxation described above is in fact not required to explain its rarity. There are overwhelmingly more microstates that correspond to a homogeneous distribution (a) than for a distribution with all particles on one side (b) \cite{huangstati}. We can therefore safely neglect the inhomogeneous macrostate and focus on the macrostate with a homogeneous  distribution of particles -- which is the very basis of statistical mechanics \cite{huangstati}.  In other words, even though there is a powerful mechanism to restore the homogeneity of the gas particles, this mechanism is not required to explain the absence of inhomogenous distributions: These macrostates are sufficiently unlikely to occur to be safely neglected. Quite similarly, as we will argue below, despite decoherence being a powerful mechanism to explain the disappearance of macroscopic quantum superpositions \cite{Zurek2003,Buchleitner:2009fk}, these states are a priori extremely rare in many ensembles of pure states.

While mixed states already include the effect of decoherence and feature little macroscopicity \cite{Frowis2012b},  we  focus here on pure states. The typicality of macroscopic entanglement then depends on the actual choice of the pure-state ensemble.  We find analytical and numerical evidence that 
 macroscopic superpositions are untypical among 
   random pure states in different ensembles. This finding may appear paradoxical at first sight, since macroscopicity leads to multipartite entanglement \cite{Morimae2010} and entanglement, as quantified by the geometric measure \cite{Streltsov2011,Shimony1995} (tantamount to a large distance to separable states), is  common in random pure states \cite{Gross2009,Bremner2009}. We resolve this  ostensible paradox by showing that geometric entanglement  is actually adverse for macroscopicity: Random states are very entangled, and \emph{hence} non-macroscopic. 

We review measures of macroscopic and geometric entanglement in Section \ref{quantification}, where we also present some technical results. Since both quantities are defined via a maximization procedure, their relationship is intricate. We elucidate their connection qualitatively in Section \ref{relationship}, in order to gain a good understanding of the statistics of macroscopic entanglement evaluated in different pure-state ensembles in Section \ref{statistics}. We conclude in Section \ref{conclusions}, where we propose an extension of our study to other ensembles and sketch its consequences for the preparation of macroscopic states in the experiment.

\section{Quantifying macroscopic and geometric entanglement} \label{quantification}
In order to address our central question -- is macroscopicity typical in ensembles of pure quantum states? -- we need to establish a quantitative measure for macroscopicity. 
Our system of interest is a collection of $N$ qubits, which, for the purpose of illustration, we treat as spin 1/2-particles. That is, we  focus on pure quantum states living in the Hilbert space $\mathcal{H}=\left( \mathbbm{C}^{2} \right)^N$.  While no consensus exists on how to rigorously quantify macroscopicity for mixed states  \cite{Dur2002,Bjork2004,Shimizu2005,Korsbakken2007,Lee2011,Frowis2012b,Xu2013}, this debate is not crucial in our context, since we focus on pure states, for which most measures agree. We will adapt a well-established measure for macroscopic entanglement \cite{Frowis2012b,Jeong2014} and propose a sensible way for its normalization. To set the context, we are interested in the typicality of macroscopic entanglement \emph{within} quantum theory; a general benchmark of macroscopic quantum superpositions will require concepts that draw beyond this realm \cite{Farrow2015,Nimmrichter2013}, and may be tailored for specific applications \cite{Laghaout2014}. 

\subsection{Measure of mascroscopicity}
Macroscopicity manifests itself in disproportionally large fluctuations of some additive multi-particle observable \cite{Frowis2012b,Frowis2014}, i.e.~of some  operator of the form
\eq
\hat S(\vec \alpha_1, \dots \vec \alpha_N) &=& \sum_{j=1}^N  \vec \alpha_j \cdot \vec \sigma_j  ,  \label{addobs}
\en 
where $\vec \sigma_j$ is the vector of three Pauli matrices that act on the $j$th qubit and the local orientation of the measurement operator $\vec \alpha_j$ is normalized, 
\eq |\vec \alpha_j |^2 =1  \label{normalizationalpha}. \en 
The operator  $\hat S$ describes the total spin of the system with respect to locally adjusted spin-directions, defined by $\vec \alpha_j = (\alpha^x_j, \alpha^y_j, \alpha^z_j)$. Given a pure quantum state $\ket{\Psi}$, the maximally obtainable variance of this additive operator, 
\eq 
\langle  \Delta \hat S^2(\vec \alpha_1, \dots \vec \alpha_N) \rangle &=& \bra \Psi \hat S^2 \ket \Psi - (\bra \Psi \hat S \ket \Psi )^2 ,   \label{variancedef}
\en
defines the unnormalized macroscopicity $\mathcal{\tilde M}$ \cite{Frowis2012b}, 
 \eq 
 \mathcal{\tilde M}= \text{max}_{\vec \alpha_1, \dots , \vec \alpha_N} \langle \Delta \hat S^2(\vec \alpha_1, \dots \vec \alpha_N) \rangle . \label{unnormalizedmacro}
 \en
Quite naturally, we find
\eq 
N \le \mathcal{\tilde M} \le N^2 .
\en
The lower bound is saturated, e.g., for separable states, for which we maximize fluctuations by choosing the measurement directions $\vec \alpha_j$ to be unbiased with respect to the local spin directions. The upper bound is reached, e.g., for the Greenberger-Horne-Zeilinger (GHZ) state  \cite{GHZori}
\eq 
\ket{\text{GHZ}}_N = \frac{1}{\sqrt 2 } \left(  \ket{0}^{\otimes N}  + \ket{1}^{\otimes N}   \right)  , \label{GHZdefinition}
\en
which comes closest to modelling a coherent superposition of a living and a dead cat: The two super-imposed alternatives (all spins pointing up, all spins pointing down) are maximally different, leading to maximal fluctuations. 

In order to compare the macroscopicity of systems of different sizes $N$, we normalize $ \mathcal{\tilde M}$:
\eq 
 \mathcal{M}(\ket{\Psi})=  \sqrt{ \frac{\mathcal{\tilde M}(\ket{\Psi})-N}{N(N-1)} },   \label{macrodef}
\en
such that  $0 \le \mathcal{M} \le 1$ for all $N$, where the upper (lower) bound is saturated for $\mathcal{\tilde M}=N^2$ $(N)$.

\subsection{Additivity and basic properties}
The unnormalized macroscopicity $\mathcal{\tilde M}$ is additive in the sense that  any product state $\ket{\Psi} \otimes \ket{\Phi}$ yields
\eq 
\mathcal{\tilde M}( \ket{\Psi} \otimes \ket{\Phi} ) = \mathcal{\tilde M}( \ket{\Psi} ) + \mathcal{\tilde M}(\ket{\Phi} )  , \label{additivity}
\en
because the separability of $\ket{\Psi} \otimes \ket{\Phi}$ excludes additional fluctuations by choosing a direction of spins other than the optimal ones for $\ket{\Psi}$ and $\ket{\Phi}$. 

Any family of states for which the size of non-separable components does not scale with the system size has vanishing normalized macroscopicity in the limit of many particles. For example, a tensor product of $N/2$ Bell-states is not macroscopic: 
\eq 
\mathcal{\tilde M} \left( \ket{\Psi^-}^{\otimes N/2} \right) &=&  2 N \\
\mathcal{M} \left( \ket{\Psi^-}^{\otimes N/2} \right)& = & \frac{1}{\sqrt{N-1}}  \stackrel{N \rightarrow \infty} {\rightarrow}  0 
\en

\subsection{Relation to index $p$}
Our measure of macroscopicity can be directly related to the index $p$ \cite{Shimizu2002},
 the exponent that defines the scaling of $\mathcal{\tilde M}$ with $N$, 
 %\eq p= \frac{\log \mathcal{\tilde M} } { \log N}  ,  \en where $p$ is
\eq 
\mathcal{\tilde M} \propto N^p .
\en
The scaling properties of macroscopic entanglement can be investigated for \emph{state families}, i.e.~``prescriptions that assign to any system size a quantum state $\ket{\Psi}$'' \cite{Frowis2014}. 
Any state family for which $\mathcal{M}>\delta >0$ with $\delta$ independent of $N$ can be considered macroscopic: In the limit $N \rightarrow \infty$, $\mathcal{M}$ is then related to the index $p$ as follows:
\eq
\mathcal{M} =0 &\Leftrightarrow& p=1 , \\
\mathcal{M} >\delta > 0 &\Leftrightarrow & p=2 .
\en
In other words, for $p=2$, we have macroscopically large fluctuations that do not vanish in the  limit of many particles; the index $p$, however, does not yield any information about the actual fraction of particles participating in a macroscopic superposition. The normalized macroscopicity $\mathcal{M}$ is more fine-grained and offers such insight: For the family of states 
 \eq 
\ket{\Psi_{(N_1,N_2)}} = \ket{\text{GHZ}}_{N_1} \otimes \ket{1}^{\otimes N_2} % \ket{ \underbrace{1, \dots , 1}_{N_2}} ,
\en
we have 
\eq
 \mathcal{\tilde M}(\ket{\Psi}) &=& N_1^2+N_2 ,
 \en
 and, thus, 
\eq
 \mathcal{M}(\ket{\Psi}) & \stackrel{ N_1, N_2 \gg 1 }{\approx }&  \frac{N_1}{N_1+N_2 } .
 \en  
In this case $\mathcal{M}$ reflects the fraction of particles in the system that take part in a quantum superposition of macroscopically distinct alternatives. On the other hand, for large $N$, we also have $N \mathcal{M} \approx \sqrt{\mathcal{\tilde M}} $ (with an additive error of the order $\sim \sqrt{N}$), which quantifies the absolute size of the macroscopic superposition. In the following, we will therefore focus on the macroscopicity $\mathcal{M}$, as defined in Eq.~(\ref{macrodef}). 

\subsection{Evaluation of macroscopicity}

\subsubsection{Generic states}
The definition of the unnormalized macroscopicity $\mathcal{\tilde M}$ in  Eq.~(\ref{unnormalizedmacro}) entails an optimization problem over $2N$ variables, which evidently complicates its evaluation for large systems.  As an alternative to  multivariable optimization, it was proposed \cite{Morimae2005} to evaluate the unnormalized macrocopicity $\mathcal{\tilde M}$ using the variance-covariance-matrix  
\eq
V_{\gamma k , \beta j} = \bra{\Psi} \Delta \hat \sigma^\gamma_{k}   \Delta \hat \sigma^\beta_{j} \ket{\Psi}  ,  \label{VCM}
\en
where $\Delta \hat \sigma^\gamma_l=  \hat \sigma^\gamma_l - \langle \hat \sigma^\gamma_l \rangle $, $\gamma=x,y,z$. The resulting $3N\times 3N$-matrix $V$ then stores the fluctuations of all observables that are sums of Pauli matrices, and we have  \cite{Morimae2005} 
\eq
\langle \Delta \hat S^2(\vec \alpha_1, \dots, \vec \alpha_N )  \rangle = \sum_{j,k=1}^N \sum_{\gamma,\beta=x,y,z} \alpha_{j}^\gamma  V_{\alpha j, \beta k} \alpha_{k}^\beta  .
 \en
 The expectation value  $\langle \Delta \hat S^2(\vec \alpha_1, \dots, \vec \alpha_N ) \rangle $ is  maximized by choosing the $\vec \alpha_j$ as the Eigenvector $\vec v_1$ of $V$ corresponding to the largest Eigenvalue $\lambda_1$ of $V$, such that 
\eq 
\vec \alpha_j = \sqrt{N}  \left( \begin{array}{c}  
v_{1,3(j-1)+1} \\
v_{1,3(j-1)+2} \\
v_{1,3(j-1)+3}   \end{array} \right)  \label{choice} 
\en
However, the $\vec \alpha_j$ chosen this way are only constrained by 
\eq 
\sum_{l=1}^N |\vec \alpha_l|^2  = N,  \label{weakcontr}
\en
which is a much weaker constraint than our Eq.~(\ref{normalizationalpha}): Instead of $N$ unit-normalized Bloch-vectors $\vec \alpha_j$,  only the sum of the norm of all Bloch-vectors is fixed in Eq.~(\ref{weakcontr}). Colloquially speaking, the $\vec \alpha_j$ chosen  according to Eq.~(\ref{choice}) \cite{Morimae2005} allow us to weight the importance of individual spins in the system differently, and give those featuring large fluctuations a larger impact.  We can therefore only state that
\eq 
\mathcal{\tilde M}(\ket{\Psi}) \le N \lambda_1 . \label{upperbound}
 \en
Admittedly, the scaling of $\lambda_1$ with the system size $N$ for a given  family of states is inherited by $\mathcal{\tilde M}$ evaluated via Eq.~(\ref{unnormalizedmacro}), such that coarse-grained quantities such as the index $p$ can be evaluated using $\lambda_1$. In particular, the Eigenvalues of $V$ only take the values 1 and 0 for  separable states, which  are clearly -- and not surprisingly -- not macroscopic at all. For a quantitative understanding of macroscopicity, however, the optimization inherent to (\ref{unnormalizedmacro}) is crucial: For example, consider the state 
\eq 
\ket{\Phi_c}=\ket{\Psi^+} \otimes \ket{0}^{\otimes N-2} , 
\en
with $N \ge 3$, where the first two qubits are in a maximally entangled Bell-state, but  remain completely separable from the rest of the system. The unnormalized macroscopicity of $\ket{\Phi_c}$ fulfils 
\eq 
\mathcal{\tilde M}= N+2 < 2 N =  \lambda_1 N ,
\en
i.e.~the largest Eigenvalue $\lambda_1$ is related to the Eigenvector $\vec v_1$, with 
\eq 
\vec \alpha_1 &=& \sqrt{\frac N 2} (1,0,0) \\
\vec \alpha_2 &=& \sqrt{\frac N 2} (1,0,0) \\
\vec \alpha_{k \ge 3} &=& (0,0,0) 
\en
which is not compatible with Eq.~(\ref{normalizationalpha}) and entirely ignores the separable qubits $3, \dots , N$, 
 while  the two entangled qubits are over-weighted. This example being admittedly artificial,  we have nevertheless experienced a substantial difference between the exact calculation and the value extracted via the VCM for the non-symmetric ensembles of random states considered below in Section \ref{statistics}.

On the other hand, given a set of orientations obtained via (\ref{choice}) and $|\vec \alpha_j|^2 >0$ for all $j$, we can normalize the $\vec \alpha_j$ to find a candidate spin orientation that promises to yield a large variance,
\eq 
\vec \beta_j = \frac{\vec \alpha_j }{ \sqrt{ | \vec \alpha_j |^2 }}  \label{betachoice} .
\en
 The resulting value $\langle \Delta \hat S^2 ( \vec \beta_1, \dots , \vec \beta_N ) \rangle$ 
 then provides a lower bound on the actual unnormalized macroscopicity, since we are not guaranteed that the choice of local spin orientations given by (\ref{betachoice}) is the optimal one: 
\eq 
\langle \Delta \hat S^2 ( \vec \beta_1, \dots , \vec \beta_N )  \rangle \le \mathcal{\tilde M}  . \label{lowerbound}
\en
In other words, even though the precise value of $\mathcal{\tilde M}$ requires a numerical optimization, computationally inexpensive lower and upper bounds [Eqs.~(\ref{lowerbound}) and (\ref{upperbound}), respectively] to this quantity can  be established straightforwardly.

\subsubsection{Symmetric states} \label{secsymmst}
In the case of permutation-symmetric states, $V$ assumes a structure with repeated $3\times 3$-blocks, and the Eigenvectors $\vec v_j$ reflect this symmetry. As a consequence, the optimal local spin orientations all coincide,  $\vec \alpha_l = \vec \alpha_k$ for all $k,l$, and, consequently, $\vec \beta_j=\vec \alpha_j$. The optimization inherent to Eq.~(\ref{unnormalizedmacro}) becomes unnecessary, since the lower bound (\ref{lowerbound}) and the upper bound (\ref{upperbound}) on the unnormalized macroscopicity coincide. We can therefore safely adopt the method introduced in Ref.~\cite{Morimae2005} to compute $\mathcal{M}$. 

Specifically, the variance-covariance-matrix $V$ then consists of two different $3\times 3$-blocks
\eq
A_{\gamma,\beta} =   V_{\gamma 1, \beta 1} , \\
B_{\gamma,\beta} =   V_{\gamma 1, \beta 2} ,
\en
which contain all variances and co-variances, respectively, and assumes the structure
\eq 
V = \left( \begin{array}{ccccc} 
 A & B &  \dots & B \\
 B& A &   \dots & B \\
 \vdots & \vdots & \ddots & \vdots \\
 B& B &  \dots & A 
\end{array}  \right) .
\en
By writing $V=\mathbbm{1} \otimes A + M \otimes B$, one can show that the largest Eigenvalue $\lambda_1$ of the above block-matrix $V$ coincides with the largest Eigenvalue of the 3$\times$3-matrix
\eq 
V_{\text{sym}} = A + (N-1) B   , \label{ABmatrix}
\en
such that 
\eq 
\mathcal{\tilde M}(\ket{\Psi_{\text{sym}}}) = N \lambda_1 . \label{nlambd1}
\en
This greatly facilitates the computation of the macroscopicity for permutation-symmetric states. To obtain the matrices $A$ and $B$, we use the efficient techniques for the computation of reduced density matrices of symmetric states presented in Ref.~\cite{Baguette2014}.

\subsection{Geometric measure of entanglement}
We will relate macroscopicity to the geometric measure of entanglement \cite{Shimony1995,Barnum2001}, which is defined via the maximal overlap $\eta$ of $\ket{\Psi}$ with any separable state $\ket{\Phi_{\text{sep}}}= \ket{\phi_1, \phi_2, \dots , \phi_N} $,
\eq 
E_{\text{G}}(\ket{\Psi} ) \equiv - \log_2 \eta = - \log_2 \sup_{\ket{\Phi_{\text{sep}}} } | \braket{\Phi_{\text{sep}}}{ \Psi } |^2   .\label{geomeasure}
\en
The geometric measure of entanglement of $N$ qubits naturally vanishes for separable states, and is bounded from above by $N-1$. High geometric entanglement is tantamount to a large generalized Schmidt measure \cite{Eisert2001}, which reflects the number of separable terms required to express the state; the geometric measure of entanglement, thus, quantifies the complexity of a quantum state. 

Since we will face the statistics of geometric entanglement in Section \ref{statistics}, we discuss its  evaluation in practice. In general, the computation of the geometric measure of entanglement requires an optimization over the $2N$ free parameters $x_1, \dots, x_N$ and $y_1, \dots , y_N$ that define the  separable state 
\eq \ket{\Phi_{\text{sep}}} = \otimes_{j=1}^N \left( \cos x_j \ket{0} + e^{i y_j} \sin x_j \ket{1} \right) , \en
 entailing significant computational expenses. Alternatively, candidate solutions for the closest separable state   can be computed via the probabilistic iterative algorithm presented in Ref.~\cite{Streltsov2011}.

For permutation-symmetric states, the evaluation of the geometric measure is facilitated  considerably. A permutation-symmetric state of $N$ qubits can be written in the Majorana-representation, 
\eq 
\ket{\Psi_{\text{sym}}}= \frac{1}{\sqrt \mathcal N} \sum_{\sigma \in S_N} \otimes_{j=1}^N \ket{\epsilon_{\sigma_j}} , \label{majorana}
\en
where 
\eq 
\mathcal{N}  = N!~ \text{perm} ( \braket{\epsilon_j}{\epsilon_k}  ) ,
\en
is a normalization  constant, and $\text{perm}( \braket{\epsilon_j}{\epsilon_k}  )$ is the permanent \cite{Minc:1984uq} of the $N\times N$ Gram-matrix that contains all mutual scalar products $\braket{\epsilon_j}{\epsilon_k} $. The permanent is, in general, a function that is exponentially hard in the matrix size $N$, but for a Gram-matrix with only two non-vanishing singular values, as given here by construction, an efficient  evaluation is possible \cite{Minc:1984uq}. For this purpose, the representation of symmetric states in the  Dicke-basis is valuable,
\eq 
\ket{\Psi_{\text{sym}}}= \sum_{j=0}^N c_j \ket{D^{(j)}_N} . \label{dickerep}
\en
The Dicke-states are defined as
\eq 
\ket{D^{(j)}_N} =  { N \choose j }^{-1/2} \sum_{\sigma \in S_{\{ 1, \dots ,1 ,0, \dots ,0 \} } }\otimes_{j=1}^N \ket{\sigma_j}  , \label{Dickedef}
\en
where the summation includes  all possibilities to distribute $j$ particles in $\ket{1}$ and $N-j$ particles in $\ket{0}$ among the $N$ modes. The  quantitative relationship between the expansion coefficients in the Dicke-basis $c_j$ and the $N$ states $\ket{\epsilon_k}$ that define the Majorana representation (\ref{majorana}) is presented in Ref.~\cite{Martin2010}. 

Since the closest separable state to a symmetric state is itself symmetric \cite{Hubener2009a}, the optimization problem over $2N$ variables implicit in (\ref{geomeasure}) reduces to a merely two-dimensional setting. Given the Majorana-representation $\ket{\epsilon_1}, \dots, \ket{\epsilon_N}$, we need to find the single-qubit state $\ket{\phi}$ that maximizes
\eq 
\frac{1}{\mathcal{N}}  \prod_{j=1}^N |\braket{\epsilon_j}{\phi} |^2. 
\en
Besides standard numerical optimization strategies, we can adapt the iterative algorithm of Ref.~\cite{Streltsov2011} to symmetric states: For that purpose, we choose a 
random single-particle state $\ket{\phi_0}$. We then iteratively generate states $\ket{\phi_k}$ as follows:
\eq 
\ket{\tilde \phi_{k+1}} &=& \sum_{j=1}^N \frac{\ket{\epsilon_j} }{\braket{\epsilon_j}{\phi_k}  } \\
\ket{\phi_{k+1}} &=& \frac{\ket{\tilde \phi_{k+1}} }{\sqrt{\braket{\tilde \phi_{k+1}}{\tilde \phi_{k+1}} }}
\en
The results of Ref.~\cite{Streltsov2011} can be translated to a wide extent to the current setting with symmetric states, and $\ket{\phi_k}$ becomes a good candidate for the closest separable state for large $k$, although the algorithm is prone to return a local instead of a global maximum.

\section{Relationship between macroscopicity and geometric entanglement} \label{relationship}
Having established the two pertinent main  characteristics of a quantum many-body state  $\ket{\Psi}$ -- its macroscopicity $\mathcal{M}$ [Eq.~(\ref{macrodef})] and its geometric entanglement $E_{\text{G}}$ [Eq.~(\ref{geomeasure})], we can now explore the relationship between these two quantities. 

The behavior of entanglement measures and macroscopicity was analysed in Ref.~\cite{Morimae2010}, where the author finds that ``a state which includes superposition of macroscopically distinct states also has large multipartite entanglement in terms of the distance-like measures of entanglement''. Here, we will confirm that geometric entanglement is indeed necessary for non-vanishing macroscopicity; however, we will also show that the general relationship between macroscopicity and geometric entanglement is rather involved. In particular, non-vanishing geometric entanglement is not sufficient for macroscopicity: There are entangled states with strictly vanishing macroscopicity. On the other hand, very large geometric entanglement implies small macroscopicity, i.e.~the maximal value of macroscopicity is reached for finite geometric entanglement. As a consequence, the two quantities should not be used synonymously.

\subsection{Close-to-separable states} \label{closesepst}
We first focus on states $\ket{\Psi}$ for which the largest squared overlap with a separable state $\eta$ is larger than or equal to 1/2, i.e.~the geometric measure of entanglement $E_{\text{G}}(\ket{\Psi})$ [Eq.~(\ref{geomeasure})] is smaller than or equal to unity. 

To this end, we explore the transition between the separable state $\ket{0, \dots, 0}$, which carries neither macroscopicity nor entanglement, to the maximally macroscopic GHZ-state (\ref{GHZdefinition}), described by 
\eq
\ket{\Xi (\theta,\epsilon) } = \frac{ \cos \theta \ket{0}^{\otimes N} + \sin \theta ( \cos \epsilon \ket{0}+ \sin \epsilon\ket{1} )^{\otimes N} }{ \sqrt {1 + \cos^N \epsilon \sin(2 \theta ) }}  ,
\en
with $0 \le \epsilon \le \pi/2$ and $0 \le \theta \le \pi/4$. The state $\ket{\Xi(\theta, \epsilon)}$ is a superposition of two separable components in which all qubits populate the very same states, with weights depending on $\theta$. The distinguishability of the two alternatives is defined by $\epsilon$.  The parametrization in $\epsilon$ for fixed $\theta=\pi/4$ was introduced  in Ref.~\cite{Dur2002} and explored in Refs.~\cite{Frowis2012b,Volkoff2014a}. Since the state is permutation-symmetric, we can use the methods of Section \ref{secsymmst} to evaluate its macroscopicity.

Several limiting cases are reached for particular values of the parameters $\theta, \epsilon$: For $(\theta, \epsilon)=(\pi/4,\pi/2)$, we recover the GHZ-state (\ref{GHZdefinition}); for $\theta=0$ or $\epsilon=0$, we deal with a separable state. For $\epsilon > \pi/2$, the destructive interference between the two amplitudes associated to the component  $\ket{0}^{\otimes N}$ can lead to geometric entanglement larger than unity. In particular, we obtain the $W$-state
\eq 
\ket{W} \equiv \ket{D^{(1)}_N} = \frac{1}{\sqrt N} \left( \ket{1, 0, \dots , 0} + \ket{0,1,0, \dots, 0} \dots  \right)  ,
\en 
in the limit $\epsilon \rightarrow \pi, \theta \rightarrow \pi/4$, for odd $N$.

We parametrize the closest separable state as 
\eq 
\ket{\Phi_{\text{sep}}(\alpha )} = \left( \cos \alpha \ket{0} + \sin \alpha \ket{1} \right)^{\otimes N} ,\label{teststate}
\en
the overlap with $\ket{\Xi(\theta, \epsilon)}$ becomes
\eq 
|\braket{\Phi_{\text{sep}}(\alpha )}{\Xi(\theta,\epsilon ) }|^2 = \frac{( \cos \theta \cos^N \alpha+ \sin \theta \cos^N(\epsilon - \alpha) )^2}{1 + \cos^N \epsilon \sin(2 \theta )} , \label{overlapas}
\en
which needs to be maximized with respect to $\alpha$ to obtain the geometric measure of entanglement. For  large $N$, the overlap is maximized for $\alpha=0$ (since  $\theta \le  \pi/4$); for finite $N$, the maximum can conveniently be found numerically, since the overlap (\ref{overlapas}) does not oscillate fast as a function of $\alpha$. 

\begin{figure}[ht]
\includegraphics[width=\linewidth]{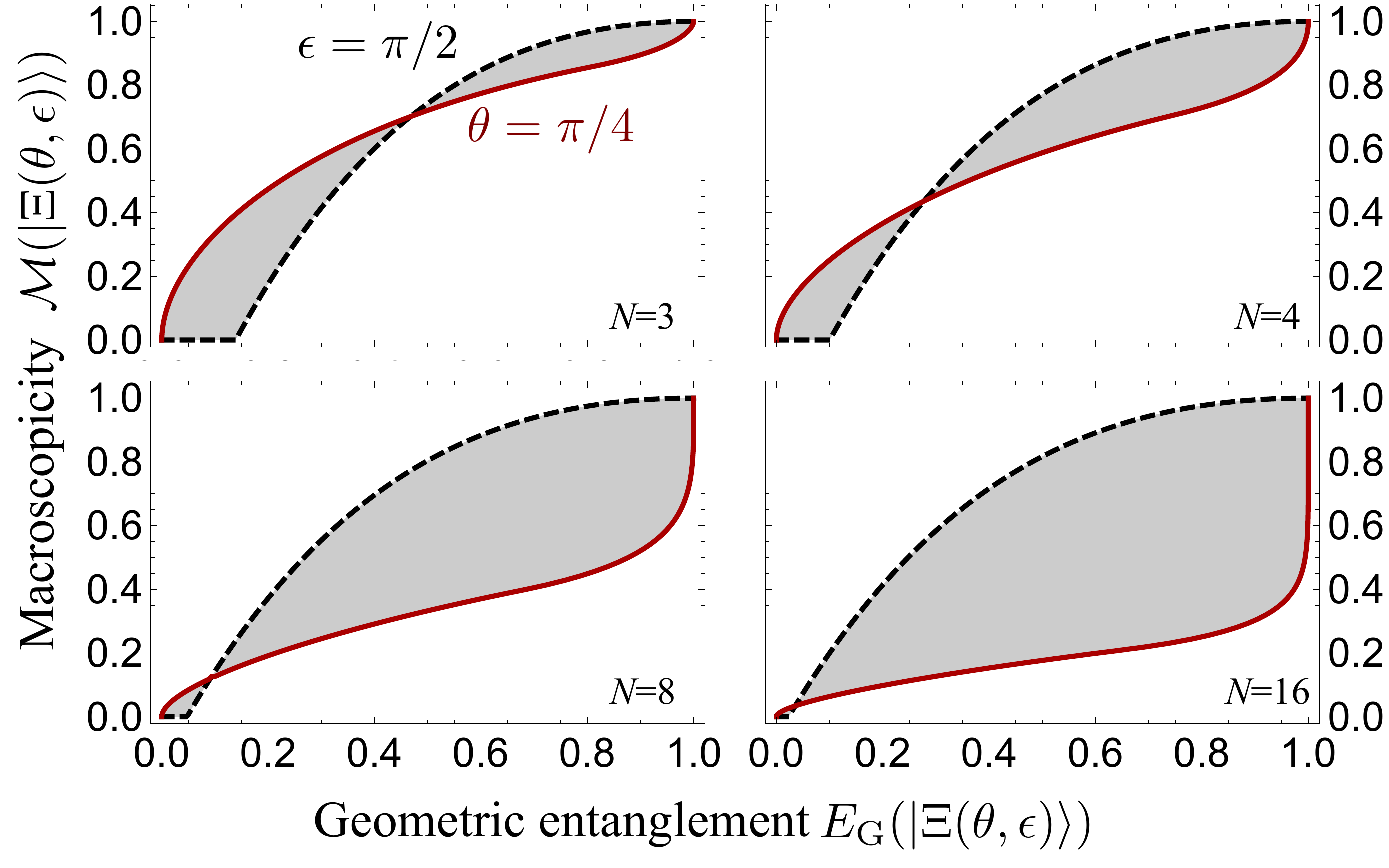}
\caption{Macroscopicity as a function of geometric entanglement for the family of states $\ket{\Xi(\theta, \epsilon)}$. The two extremal cases are given by $\epsilon=\pi/2$ (black dashed) and $\theta=\pi/4$ (solid red),  which  take turns as the upper bound on  macroscopicity for given geometric entanglement. }
\label{overviewXistate.pdf}
\end{figure}

We show the behavior of geometric and macroscopic entanglement in Fig.~\ref{overviewXistate.pdf}, for different numbers of qubits $N$. Although macroscopicity increases with geometric entanglement as a general tendency, the relationship is ambiguous, especially for large numbers of qubits $N$. Based on extensive numerical evidence, we conjecture that the general maximum macroscopicity for a given value of geometric entanglement is attained by the value obtained for $\ket{\Xi(\theta = \pi/4, \epsilon)}$ or $\ket{\Xi(\theta , \epsilon=\pi/2) }$, in the range $E_{\text{G}} \le 1$. 

\subsubsection{Geometric entanglement without macroscopicity} \label{geomwim}
We first consider the family of states parametrized by $\epsilon=\pi/2, 0 \le \theta \le \pi/4$ (black dashed lines in Fig.~\ref{overviewXistate.pdf}), the pertinent variance-covariance matrix $V_{\text{sym}}$ in Eq.~(\ref{ABmatrix}) then becomes
\eq 
V_{\text{sym}}     = \left( \begin{array}{ccc} 1 & 0 & 0 \\ 0 & 1 & 0 \\ 0 & 0 & N \sin^2 2 \theta \end{array} \right) ,
\en
which yields the macroscopicity, 
\eq 
\mathcal{M} =\sqrt{\frac{ \text{max} ( 1, N \sin^2 2 \theta ) -1 }{N-1} }. 
\en
The geometric entanglement takes the value
\eq 
E_{\text{G}} = - \log_2 ( \cos^2 \theta   ) .
\en
As a consequence, for $0 < \theta  \le \frac 1 2 \arcsin \sqrt{1/N} $, the geometric entanglement remains  finite, yet the macroscopicity vanishes.  In other words, there are states that are entangled, but the fluctuations in any additive observable 
 do not surpass those that can be achieved for separable states. 
  This explains the step-like behavior observed in Fig.~\ref{overviewXistate.pdf}, best visible for small $N$.  On the other hand, even though geometric entanglement is not sufficient for macroscopicity, it is necessary: For a separable state, all non-vanishing Eigenvalues of the variance-covariance matrix (\ref{VCM}) are unity. 

\begin{figure}[ht]
\includegraphics[width=\linewidth]{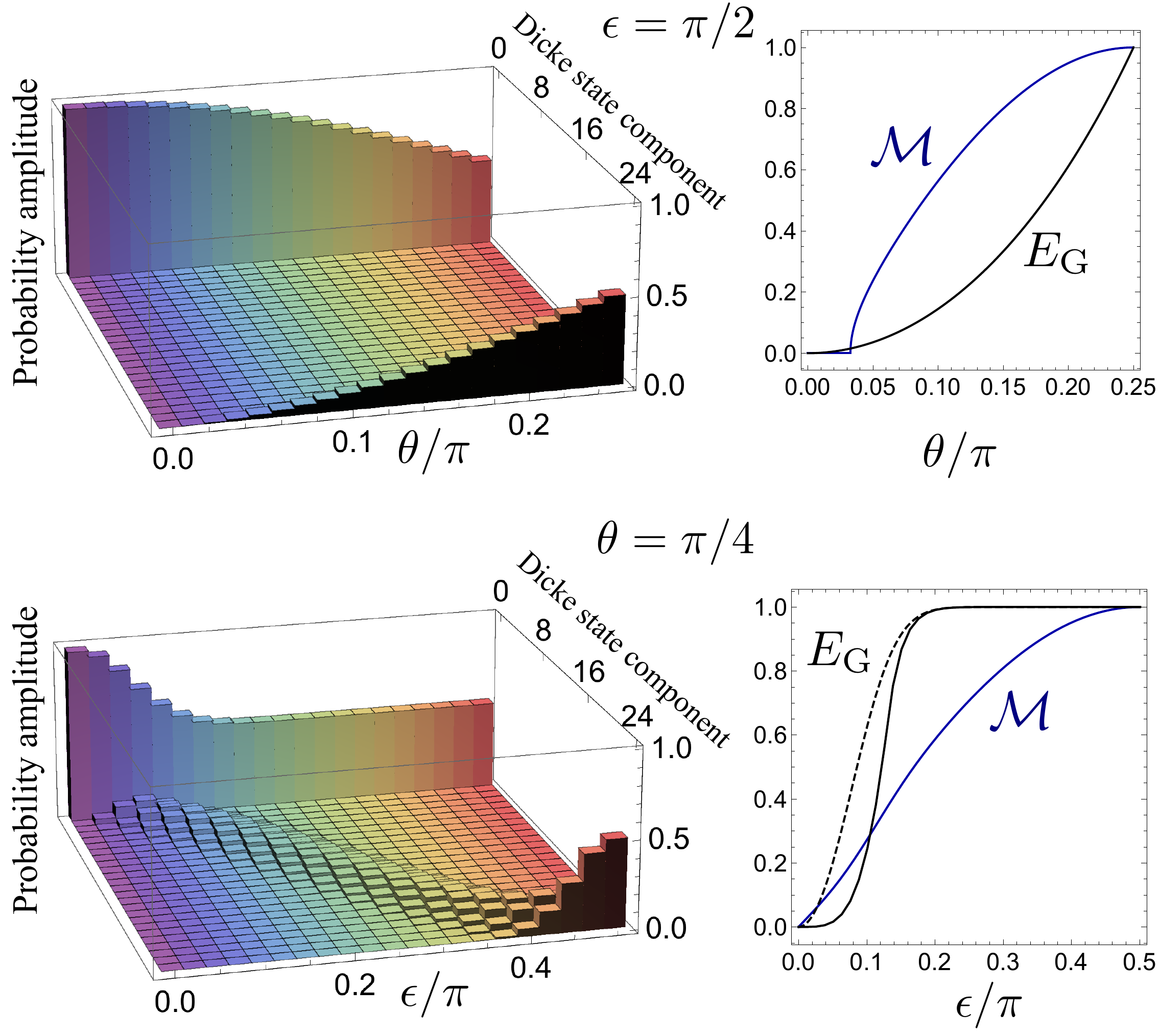}
\caption{Decomposition of $\ket{\Xi(\theta, \epsilon)}$ for $N=24$ into Dicke-state-components [Eq.~(\ref{Dickedef})], for the two extremal families of states characterized by $\epsilon=\pi/2$ (upper panels) and $\theta=\pi/4$ (lower panels). For $\epsilon=\pi/2$, only the very first and very last Dicke-components are populated, while a binomial distribution of components slowly shifts to the highest Dicke-component for $\theta=\pi/4$. Consequently, the behavior in the $E_{\text{G}}- \mathcal{M}$-plane [Fig.~\ref{overviewXistate.pdf}] is very different for the two families of states. In the lower right panel, the black dashed line shows the upper bound to $E_{\text{G}}$ given in Eq.~(\ref{upperboundgoe}).}
\label{transitionvisu.pdf}
\end{figure}

\subsubsection{Close-to-unity geometric entanglement and small macroscopicity}
The most macroscopic state, the GHZ state (\ref{GHZdefinition}), possesses geometric entanglement $E_{\text{G}}(\ket{\Psi_{\text{GHZ}}})=1$. Anticipating Section \ref{ffss}, large macroscopicity naturally comes with geometric entanglement $E_{\text{G}} \approx 1$, in general. The criterion $E_{G} \approx 1$ is, however, not sufficient to ensure large macroscopicity: Consider the family of states parametrized by $\theta=\pi/4, 0 \le \epsilon \le \pi/2$. By evaluating the largest Eigenvalue of the matrix $V_{\text{sym}}$, we obtain the macroscopicity 
\eq 
\mathcal{M} & = & \sqrt{\frac{ \sin^2 \epsilon } { 1 +\cos^N \epsilon} } .
\en
The maximal overlap to separable states is bounded from below by the overlap with the  separable test-state (\ref{teststate}) setting  $\alpha=0$; the geometric measure of entanglement therefore fulfils 
\eq 
E_{\text{G}}(\ket{\Xi(\theta=\pi/4, \epsilon)}) \le  - \log_2 \frac{ (1+\cos^N \epsilon  )}{2} . \label{upperboundgoe}
\en
That is, for a wide range of $\epsilon \leq \pi/2 $, we retain a geometric measure of entanglement of around unity, but quickly loose macroscopicity. In Fig.~\ref{overviewXistate.pdf}, the red lines quickly dive into low values of macroscopicity,  while remaining close to $E_{\text{G}} =1$, a trend that becomes more and more clear for larger values of $N$. This stands in stark contrast to the black dashed lines, which retain macroscopicity for decaying geometric entanglement. 

This behavior can be understood intuitively via the decomposition of the state into Dicke-states [Eq.~(\ref{dickerep})], shown in Fig.~\ref{transitionvisu.pdf}. The largest overlap with any separable state is at least as large as the coefficient in the Dicke-state expansion related to $\ket{D^{(0)}_N}$, since the latter is separable. For decreasing $\epsilon \approx \pi/2$, we continuously loose macroscopicity, because the average directions of the spins become similar and the two superimposed alternatives less and less macroscopically distinct. However, the closest separable state remains the Dicke-state $\ket{D^{(0)}_N}$ for a wide range of $\epsilon \leq \pi/4$, i.e.~the geometric measure of entanglement remains close to unity. In contrast, for the family parametrized by $\epsilon=\pi/2$, the loss of geometric entanglement is directly accompanied by a loss of macroscopicity (upper panel of Fig.~\ref{transitionvisu.pdf}).

\subsection{Far-from-separable states}  \label{ffss}
Let us now move into the domain of strongly geometrically entangled states and assume that we are given a maximal overlap with separable states $\eta < 1/2$, i.e.~a geometric measure of entanglement $E_{\text{G}}= -\log_2 \eta > 1$.  We construct a state $\ket{\Psi_\eta}$ that maximizes the macroscopicity under this constraint. Since local rotations do not affect the geometric measure of entanglement, we can assume that the optimal value of the spin-orientations all point into the $z$-direction ($\alpha_j=(0,0,1)$), and we expand 
$\ket{\Psi}$ in Eigenstates of the total spin operator $\hat S$, which has  
 Eigenvalues $-N, -N+2, \dots , N$. Each Eigenvalue is ${N \choose (N+S)/2 }$-fold degenerate, the state can therefore be written as 
\eq 
\ket{\Psi_\eta} = \sum_{S=-N, -N+2, \dots N} \sum_{\lambda=1}^{{N \choose (S+N)/2 }}  c_{S,\lambda} \ket{S,\lambda} , \label{expansionspin}
\en
where $\lambda$ labels the degenerate states, and  we choose the $\ket{S,\lambda}$ to be separable. The expectation value of powers of the collective spin becomes
\eq 
\bra{\Psi} \hat S^k \ket{\Psi} & =&  \sum_{S=-N, -N+2, \dots N} S^k \sum_{\lambda=1}^{{N \choose (S+N)/2 }}   |c_{S,\lambda}|^2  .
\en
Under the constraint that the maximal overlap to any separable state be fixed to $\eta$,
\eq 
|c_{S,\lambda}|^2 \le  \eta , 
\en 
 we maximize the variance (\ref{variancedef}) 
by setting $c_{N,1}=c_{-N,1}=\sqrt{\eta}$ and subsequently distributing the probability amplitude $\sqrt{1- 2 \eta}$ among the remaining coefficients, i.e.~we set as many pairs of coefficients  to $c_{S,k}=c_{-S,k}=\sqrt{\eta}$ as possible, proceeding from large to small total spins $S$. That is, we maximize the  contribution to the expectation value of $S^2$, while the expectation value of $S$ remains 0. The last pair $c_{S,k}=c_{-S,k}$ is set to accommodate the remaining amplitude, typically smaller than $\sqrt{\eta}$. Formally, the state  reads
\eq 
\ket{\Psi_{\eta,\text{max}}} =  \sum_{S=-N,-N+2, \dots , -\text{mod}(N,2)  } \times \nonumber ~~~~~~~~~ \\ 
~~~~~~~ \sum_{k=1}^{ {N \choose (N-S)/2  } } \sqrt{\eta} \left( e^{i \phi_{S,k}} \ket{S,k} +   e^{i \phi_{-S,k}}  \ket{-S,k}  \right) ,
\en
where the sum only runs over so many terms such that the state is normalized to unity, one term may possibly be weighted by a factor smaller than $\sqrt{\eta}$. 

\begin{figure}[ht]
\includegraphics[width=.8\linewidth]{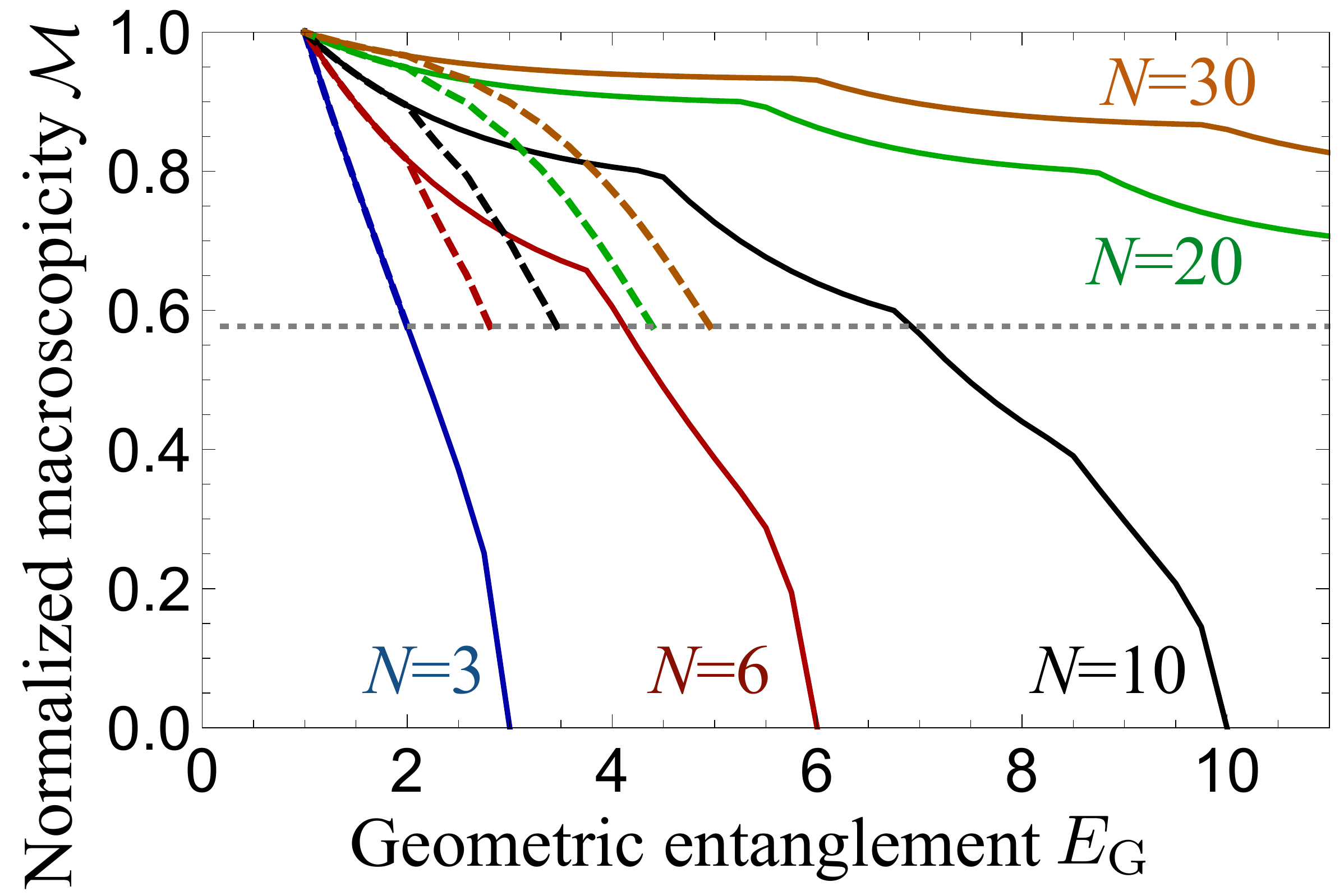}
\caption{Maximal value of normalized macroscopicity, evaluated using the argument of Section \ref{ffss}, for general states (solid lines) and symmetric states (dashed lines), for $N=3,6,10,20,30$ (blue, red, black, green and orange, respectively). The horizontal dotted line indicates the limiting value $1/\sqrt{3}$ for symmetric states (Section \ref{ransymst}). For $N=3$, the bounds for general and for symmetric states coincide. 
}
\label{maxima.pdf}
\end{figure}

The obtained bound is shown in Fig.~\ref{maxima.pdf} for different values of $N$, as a function of geometric entanglement.  The maximally achievable macroscopicity grows as a function of $N$ for a fixed overlap $\eta$, but, for a fixed number of qubits $N$, a small overlap $\eta$, equivalent to large geometric entanglement, causes a reduced macroscopicity. The expansion into separable states Eq.~(\ref{expansionspin}) is, however, not necessarily the optimal generalized Schmidt decomposition \cite{Eisert2001}, i.e.~it is often possible to find a separable state with overlap larger than $\eta$.  We can therefore not expect the bounds to be tight.

We can repeat the argument for symmetric states, for which we impose that the amplitude of each Dicke-state component $\ket{D^{(k)}_N}$ is constrained by $\eta$, leading to a state of the form 
\eq 
\ket{\Psi_{\eta,\text{max}}} =  \sum_{k=0 \dots  \lfloor N/2 \rfloor} \sqrt{\eta} \left[ e^{i \phi_{k}} \ket{D^{(k)}_{N}} +   e^{i \phi_{N-k}}  \ket{D^{(N-k)}_{N}} \right] .
\en
Due to the strict symmetry constraint on the state, we obtain smaller maximal values of macroscopicity for given geometric entanglement (dashed lines in Fig.~(\ref{maxima.pdf})) than for general states. 

\section{Statistics of macroscopic and geometric entanglement} \label{statistics}

Having established the relationship between geometric and macroscopic entanglement, we proceed to numerical investigations of pure states in different ensembles. 

\subsection{Random physical states} \label{randphysstse}

\subsubsection{State generation}
As a first ensemble of pure states, we consider random physical states, introduced in Ref.~\cite{Hamma2012a,Hamma2012} and sketched in Fig.~\ref{sketchrandomphysical.pdf}(a). To generate a random physical state $\ket{\Psi_k}$,  we assume that the qubits are aligned in  spin-chain configuration and that only pairwise interactions take place. We apply $k$ times a random two-particle unitary onto a randomly chosen pair of two neighboring qubits. That is, for $k< N-1$, the state remains at least $1$-separable (the first qubit in the chain has never directly or indirectly interacted with the  last one), while we obtain Haar-random states in the limit $k \gg N$, which we will discuss separately in Section \ref{haarrandomstates} below. To obtain a better intuition for this ensemble, we compare random physical states to random linear chains $\ket{\Phi_k}$, which are generated by applying a binary unitary between the first $k \le N-1$ pairs of qubits, starting from a separable state [Fig.~\ref{sketchrandomphysical.pdf}(b)]. As the authors of \cite{Hamma2012a,Hamma2012} argue, the ensemble of random physical states can be regarded as typical for physical systems that obey some locality structure.  A variant of random physical states for which the binary interactions are chosen to be not necessarily between adjacent neighbors but between any two randomly chosen qubits does not exhibit qualitative differences to the locality-preserving model here. 

\subsubsection{Numerical results}

\begin{figure}[ht]
\includegraphics[width=\linewidth]{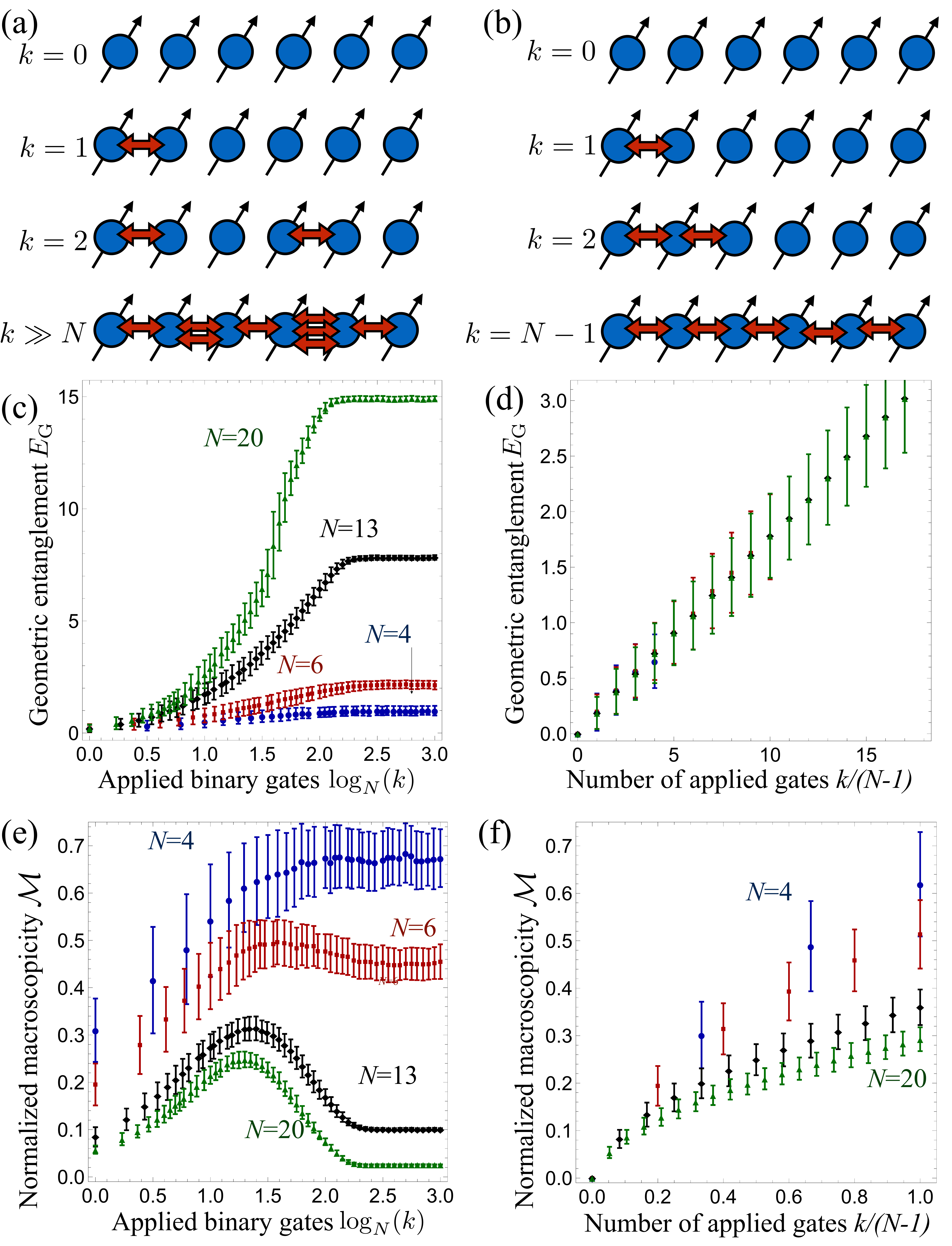}
\caption{(a) Random physical states. We apply $k$ times a random unitary between two randomly chosen adjacent sites (closed boundary conditions). For $k \gg N$, we have a fully connected system with  high probability, i.e.~the state is typically 0-separable, for $k \rightarrow \infty$, we reach the limit of Haar-uniform states. (b) Random linear chain. We apply random unitary binary gates between the first $k\le N-1$ pairs of adjacent qubits. For $k=N-1$, we have a fully connected system. Geometric entanglement (c,d) and normalized macroscopicity (e,f) for random physical states and random linear chains, for $k$ random two-body unitaries, respectively. In (d), the geometric entanglement coincides for all  numbers of qubits $N$. To compare different system sizes, we plot the entanglement and the normalized macroscopicity as a function of the $N$-base logarithm of $k$ in (c,e), and as a function of the normalized number of applied gates in (d,f). Sample size is 200, error bars show one standard deviation. }
\label{sketchrandomphysical.pdf}
\end{figure}

\begin{figure}[ht]
\includegraphics[width=.75\linewidth]{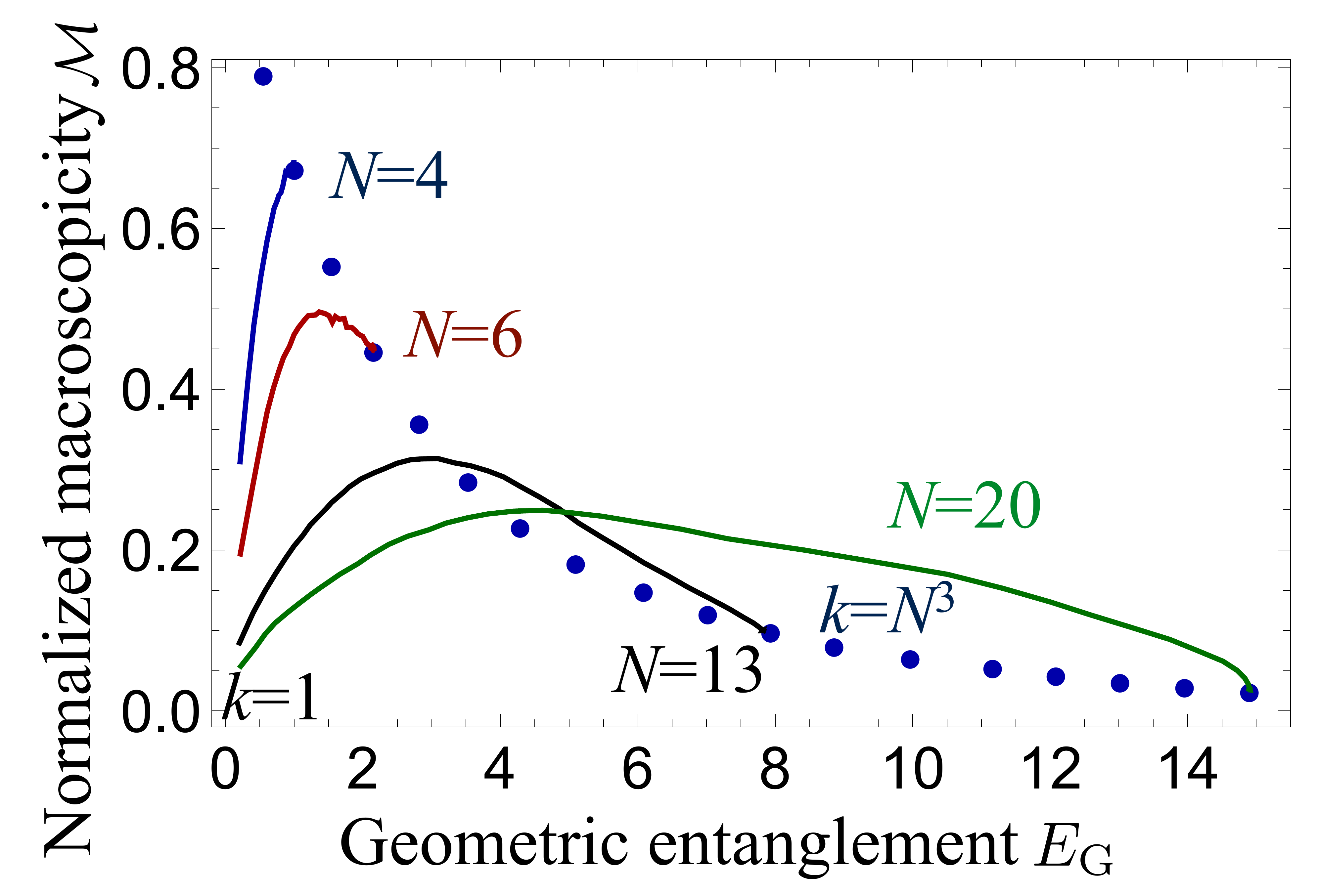}
\caption{Average trajectories of random physical states $\ket{\Psi_k}$ in the $E_{\text{G}} - \mathcal{M}$-plane for $N=4,6,13,20$. The solid lines start at $k=1$ and proceed to $k=N^3$ (blue discs). }
\label{trajectory.pdf}
\end{figure}

Geometric and macroscopic entanglement for  random physical states are shown in Fig.~\ref{sketchrandomphysical.pdf}(c,e) as a function of the number of applied binary gates $k$, which can be confronted to the behavior of random linear chains [Fig.~\ref{sketchrandomphysical.pdf}(d,f)]. 
Between $k \approx N$ and $k \approx N^2$, we observe a steep increase in the geometric entanglement of random physical states: For this range of numbers of binary interactions $k$, the state typically becomes fully inseparable. For $k \approx N^3$, we observe a saturation of  both macroscopicity and geometric entanglement. 
 While geometric entanglement increases monotonically with the number of applied gates, macroscopicity develops a peaked structure for $N \geq 6$. Consequently, the trajectories of random physical states as a function of $k$ in the $(\mathcal{M},E_{\text{G}})$-plane [Fig.~\ref{trajectory.pdf}] proceed from  low geometric and macroscopic entanglement ($k=1$) over a maximum to the asymptotic value with large geometric and low macroscopic entanglement. The maximum value of macroscopicity is reached for $ N \leq k \leq  N^2$: In this range, the states can very probably not be decomposed into separable components, while it remains moderately complex by construction -- these are the very requirements for high macroscopicity. Random linear chains feature a linear increase of geometric entanglement with the number of applied gates, for which the curves for all particle numbers coincide [Fig.~\ref{sketchrandomphysical.pdf}(d)], and a monotonic increase of macroscopicity for $k$, peaking at lower and lower values as we increase the number of particles [Fig.~\ref{sketchrandomphysical.pdf}(f)].

\begin{figure}
\includegraphics[width=\linewidth]{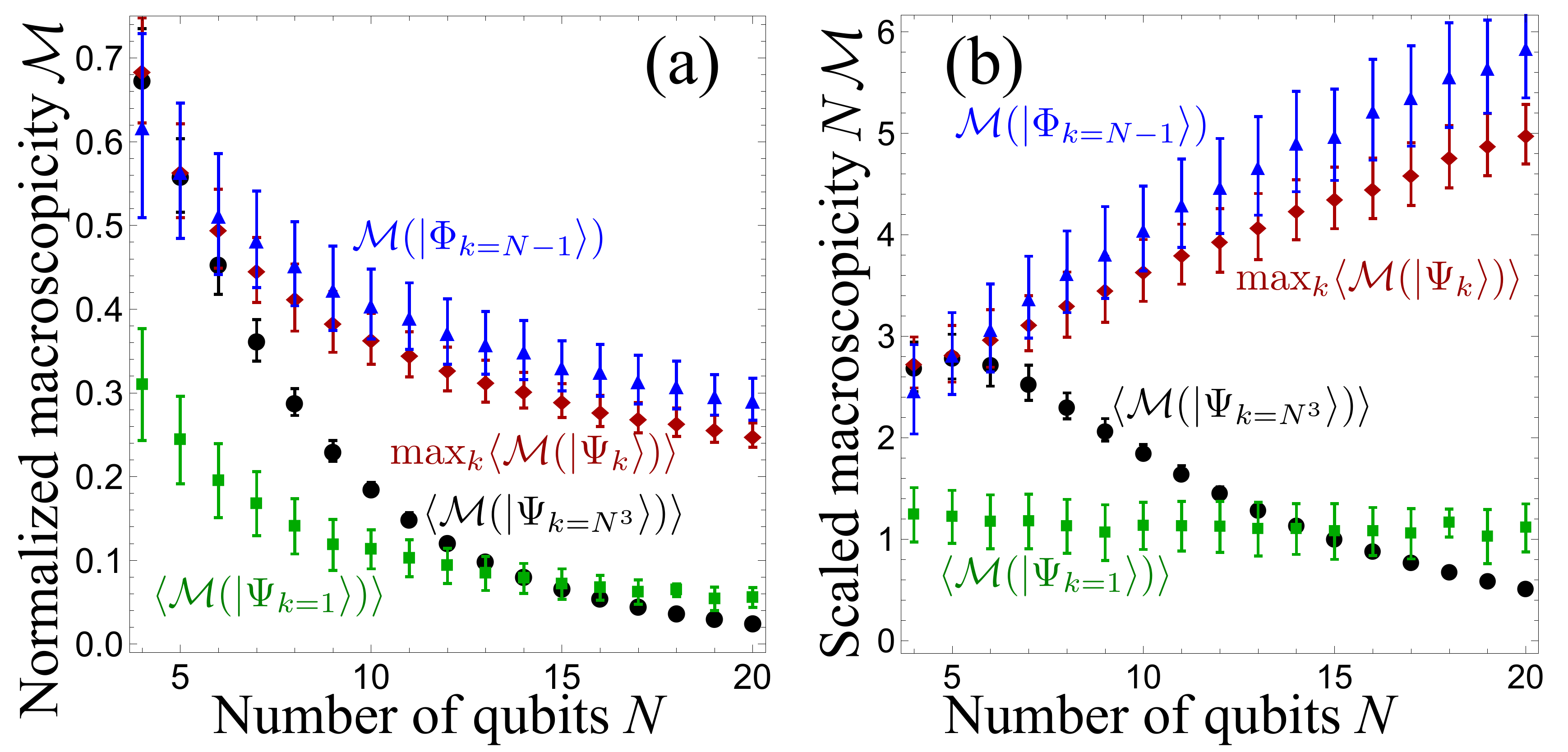}
\caption{
(a) Normalized macroscopicity $\mathcal{M}$ after one random gate ($k=1$, green squares), after $k=N^3$ random gates (black discs), maximal attained macroscopicity for random physical states (red diamonds) and normalized macroscopicity for a binary chain with $k=N-1$ random interactions (blue triangles), as a function of the number of qubits $N$. (b) Scaled macroscopicity $N \mathcal{M} \approx \mathcal{\tilde M}$ for the same ensembles of states as in (a). Error bars show one standard deviation.} 
\label{ranphys2.pdf}
\end{figure}

Macroscopicity as a function of the particle number $N$ is plotted in Fig.~\ref{ranphys2.pdf}. The maximum value in random physical states decreases with increasing $N$ (red diamonds), albeit slower than the saturated value of the macroscopicity ($k=N^3$, black circles). The latter remains slightly lower than the macroscopicity reached for a saturated random linear chain (i.e.~after $k=N-1$ binary gates) -- choosing the  interacting qubits randomly is disadvantageous for large macroscopicity, which gives an advantage to saturated linear chains. Complexity is adverse to macroscopicity: For $N \ge 15$, one randomly chosen binary interaction onto two qubits in an initially separable state $(k=1$, green squares) results in a larger macroscopicity than the limiting case $k \rightarrow N^3$.  

The different types of decay raise the question whether, albeit the fraction of particles participating in macroscopic superpositions decreases, the absolute number may in fact be constant or increase. We plot the absolute size of the macroscopic component  $N \mathcal{M}\approx \mathcal{\sqrt{\tilde M}}$ -- the approximation is justified for $N \gg 1$ --, which resolves the qualitative differences between the ensembles: The absolute number of particles participating in a GHZ-like state remains constant for states into which exactly one random gate has been applied (green squares), it decreases for $k=N^3$ (black circles), but it increases with $N$ for the maximally achieved value in random physical states and for saturated random linear chains ($k=N-1$).  

In conclusion, starting from a separable state and applying random binary gates, we first explore the region in which geometric and macroscopic entanglement are synonymous (Section \ref{closesepst}), such that both quantities initially grow with $k$. When the spin-chain is fully inseparable, additional interactions contribute to larger geometric entanglement, but simultaneously destroy its macroscopicity ensuring that the latter decreases  (Section \ref{ffss}). Even thought the absolute size of the macroscopic component increases with $N$ [Fig.~\ref{ranphys2.pdf}(b)], the fraction of particles participating in a macroscopic superposition does not. To come back to our classical analogy [Fig.~\ref{analogy}], just like there are no concerted forces that spontaneously push all gas particles to one side of the box, random evolutions are unlikely to force all spins into a macroscopic superposition.

\subsection{Haar-random states} \label{haarrandomstates}

\subsubsection{State generation}
In the limit $k \rightarrow \infty$, random physical states converge to Haar-random states, i.e.~the ensemble of pure quantum states that are uniformly distributed on the unit sphere in Hilbert space \cite{Wootters1990}. Instead of applying many binary gates, one can construct Haar-random states by randomly generating the real and imaginary part of each state coefficient $c_{j_1, \dots , j_{2^N}}$ following a zero-mean unit-variance normal distribution, the resulting unnormalized vector $\vec c$ is then normalized in a second step. This procedure yields a ``chaotic'' ensemble \cite{Sugita2005} that remains invariant under local basis-rotations and  re-partitioning of the Hilbert-space into subsystems \cite{Tichy2011a}. Random states also result from the application of a Haar-random unitary matrix on any constant pure state \cite{Zyczkowski1994}. 

\subsubsection{Macroscopicity is rare in Haar-random states}
Random states feature the {concentration of measure} phenomenon, i.e.~most states on the high-dimensional Bloch-sphere lie close to the equator \cite{DissTiersch}. Given a  Lipschitz-continuous function $f(\ket{\Psi})$, the function values remain close to the average value $\langle f \rangle$ for the vast majority of states, reflected by the probability for a deviation larger than $\epsilon$ \cite{ledoux},
\eq 
P [| f( \ket{\Psi}) - \langle f \rangle  | > \epsilon ] \le 4 e^{-\frac{(n+1) \epsilon^2}{24 \pi^2 \eta^2}} ,
\en
where 
 $\eta$ is the Lipschitz constant. Using trial functions, 
 we find that the macroscopicity defined in Eq.~(\ref{macrodef}) is Lipschitz-continuous, while 
  the geometric measure of entanglement inherits Lipschitz-continuity from the distance-like measure it is based on \cite{DissTiersch}. Hence, most Haar-random states are very similar, both when characterized by their geometric entanglement and by their macroscopicity. 

Using random matrix theory \cite{Ullah1964}, one can estimate the typical magnitudes of the elements of the variance-covariance-matrix (\ref{VCM}) \cite{Sugita2005}. In the limit  $N \rightarrow \infty$, the VCM approaches the unit matrix and, as the largest Eigenvalues converge to unity, by the upper bound (\ref{upperbound}), the normalized macroscopicity vanishes. This result also follows from the following complementary argument: Random states that are chosen according to the Haar measure  possess large geometric entanglement \cite{Bremner2009,Gross2009}: With probability greater than $1-e^{-N^2}$, we have for $N \ge 11$ \cite{Gross2009}
\eq 
E_{\text{G}}(\ket{\Psi_{\text{random}}}) \ge  N - 2 \log_2 N -3 \label{grossbound} .
\en 
 The upper bound on macroscopicity in Section \ref{ffss} then implies that the typical macroscopicity is necessarily small: Strongly geometrically entangled states cannot be macroscopic. 
 
\begin{figure}[ht]
\includegraphics[width=\linewidth]{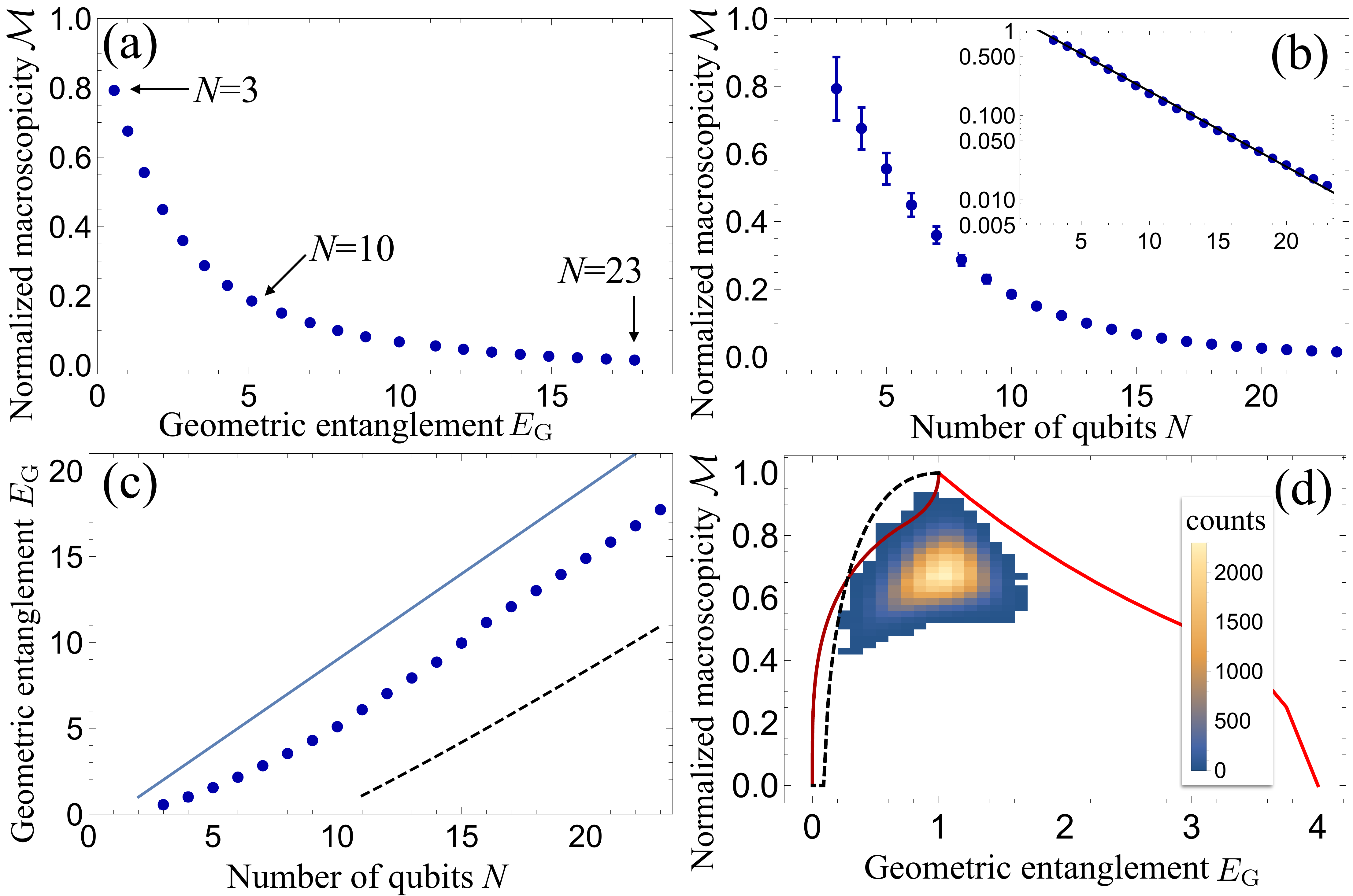}
\caption{
Macroscopicity of Haar-random states. Error-bars show one standard deviation. (a) Average macroscopicity and geometric entanglement, for $N=3, \dots, 23$, no error bars shown.  (b) Macroscopicity as a function of the number of qubits. Inset: Logarithmic plot with fit by an exponential decay. (c) Geometric entanglement. Blue solid line: Maximally possible value of geometric entanglement $E_{\text{G}}=N-1$. Dashed black line: Lower bound on geometric entanglement of random states, Eq.~(\ref{grossbound}). Error bars are not visible. Sample sizes (a-c) $N=3-10: 10^5, N=11-15: 10^4, N=16-23: 10^3$. (d) Histogram for $10^5$ Haar-random states of $N=4$ qubits, together with the upper bound of Section \ref{ffss} (red solid line, $E_{\text{G}}\ge 1$) and the state $\ket{\Xi}$ parametrized as in Fig.~\ref{overviewXistate.pdf} (red solid line and black dashed line, $E_{\text{G}} \le 1$.  }
\label{haarrandomdata.pdf}
\end{figure}

 \subsubsection{Numerical results}
The expected behavior is reproduced by our numerical data, shown in Fig.~\ref{haarrandomdata.pdf}: In agreement with the previous argument, the geometric entanglement increases as a function of the number of qubits $N$ (c), while the normalized macroscopicity decays (b). This decay is approximately exponential [logarithmic inset of Fig.~\ref{haarrandomdata.pdf}(b)], and we can safely state that  even the unnormalized macroscopicity $\mathcal{\tilde M}$ (Eq.~\ref{unnormalizedmacro}), which reflects the absolute size of the macroscopic component,  decreases. The variances of both normalized macroscopicity and geometric entanglement decrease as well. The histogram in Fig.~\ref{haarrandomdata.pdf}(d) shows the distribution of states for $N=4$ in the $(\mathcal{M}, E_{\text{G}})$-plane, together with the bounds in the regime $E_{\text{G}}\ge 1$ and the conjectured bounds in the realm  $E_{\text{G}} \le 1$.

In conclusion, both the relative and the absolute size of the largest macroscopic superposition in Haar-random states decreases with $N$. As a consequence, Haar-random states are very geometrically entangled and feature little macroscopicity. 

\subsection{Random symmetric states} \label{ransymst}
Colloquially speaking, Haar-random states are extremely complex and do not allow any efficient description \cite{Venzl2009}. A state with large macroscopicity, on the other hand, can be approximated by a superposition of Eigenstates of the total spin operator [Eq.~(\ref{expansionspin})], and thereby permits  an efficient description. Hence, complexity and macroscopicity are mutually exclusive properties, and we cannot expect to encounter macroscopic superpositions in structureless ensembles. 

On the other hand, ensembles of random pure states that are less complex may feature higher values of macroscopicity. In particular, permutationally symmetric states constitute an ensemble of states with rather low geometric entanglement \cite{Markham2011,Aulbach2010,Martin2010,Baguette2014}: 
\eq 
E_{\text{G}}(\ket{\Psi_{\text{sym}}}) \le \log_2(N+1)  \label{boundsymmetric} ,
\en
due to the vastly reduced dimensionality $N+1$ of the space of symmetric $N$-qubit-states in contrast to the full Hilbert-space of size $2^N$. We choose the following ensemble of symmetric states: In the Dicke-state representation [Eq.~(\ref{dickerep})], the coefficients $c_{j}$ are chosen to be normally distributed random variables (with normal real and imaginary parts), and the resulting states are normalized. Following this prescription, the ensemble is invariant under permutation-symmetric local unitary operations. 

\begin{figure}[th]
\includegraphics[width=\linewidth]{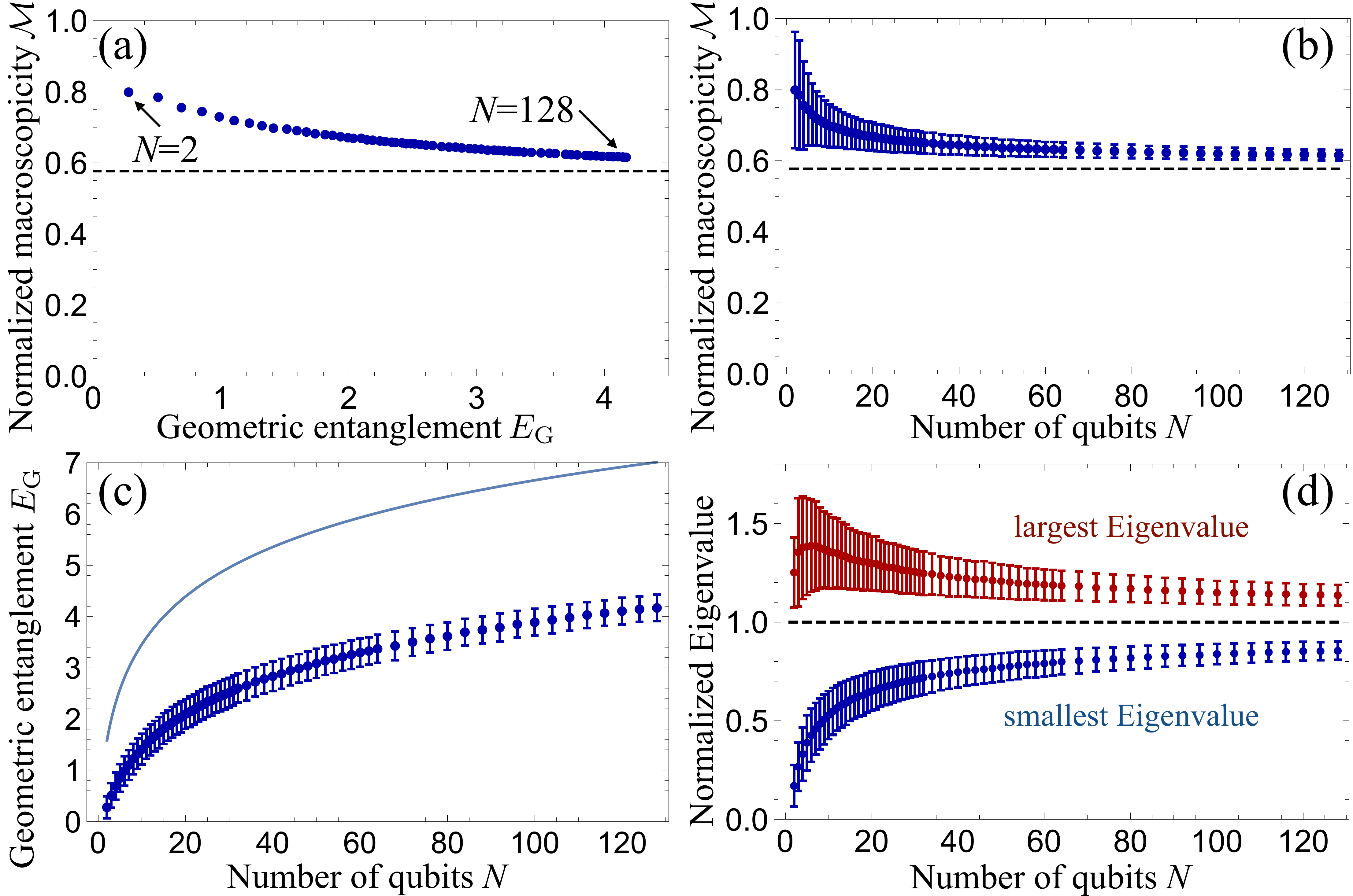}
\caption{
Macroscopicity of random symmetric states for a sample of 3000 random states. Error-bars show one standard deviation. (a) Normalized macroscopicity against geometric entanglement, for $N=2,\dots, 128$.  (b) Average normalized macroscopicity as a function of the number of qubits $N$, the dashed black line shows the limiting value $1/\sqrt 3$. 
(c) Average geometric entanglement as a function of $N$, the solid line shows the theoretical maximum $\log_2 (N+1)$ [Eq.~(\ref{boundsymmetric})].
(d) Average normalized largest and smallest Eigenvalues $\lambda / (1+(N-1)/3)$  of the matrix $V_{\text{sym}}$ for randomly chosen symmetric states. The largest Eigenvalue is directly related to the macroscopicity via Eq.~(\ref{nlambd1}). The smallest Eigenvalue solves the minimization problem that consists in  finding the additive observable with the weakest fluctuations. For large $N$, the local spin orientation is rather irrelevant for experiencing large fluctuations, as long as all local spin measurements are performed along the same axis.}
\label{EigenvaluePlot.pdf}
\end{figure}

The very different behavior of geometric entanglement for Haar-random and random symmetric states is evident comparing Figs.~\ref{haarrandomdata.pdf}(c) and \ref{EigenvaluePlot.pdf}(c).  Consistent with their low geometric entanglement, symmetric states feature exceptionally high and robust macroscopicity. 

Expectation values of observables read 
\eq
\bra{\Psi_s} \hat \sigma_k \otimes \hat \sigma_l \ket{\Psi_s} = \sum_{p,q=0}^N c_p^* c_q \bra{D^{(p)}_N} \hat  \sigma_k \otimes \hat \sigma_l \ket{D^{(q)}_N}  .
\en
Since the $c_p$ are chosen randomly and independently, only the summands with $p=q$ will contribute to the average in the limit of many qubits $N \rightarrow \infty$. 
 The only non-trivial expectation values that do not vanish on average are the two-qubit-correlations along the same axis, 
\eq
\bra{\Psi_s} \hat \sigma_k \otimes \hat \sigma_k \ket{\Psi_s} = \frac 1 3 .
\en
Consequently, the matrix $V_{\text{sym}} $ in Eq.~(\ref{ABmatrix}) converges to 
\eq 
V_{\text{sym}} = \left(1+\frac{ N-1}{3} \right) \mathbbm{1} ,
\en
with obvious Eigenvalues, and, in the limit $N\rightarrow \infty$, we therefore expect that the macroscopicity approaches
\eq 
\mathcal{M}(\ket{\Psi_s}) \rightarrow \frac{1}{\sqrt 3} . \label{sqrt3}
\en
For finite $N$, off-diagonal non-vanishing correlations may contribute further to the fluctuations, which is why the average macroscopicity converges to $1/\sqrt{3}$  from above. This behavior is confirmed empirically in Fig.~\ref{EigenvaluePlot.pdf}(a,b), where the average value of macroscopicity for symmetric states is plotted against the average geometric entanglement (a) and the number of qubits (b). 

Moreover, not only does the macroscopicity converge to a finite value, it is also very robust with respect to mis-alignment of spin-orientations: Since the smallest and largest Eigenvalues of  $V_{\text{sym}}$ [Eq.~(\ref{ABmatrix})] converge to the same value $1+(N-1)/3$ [Fig.~\ref{EigenvaluePlot.pdf}(d)], the spin-orientation becomes irrelevant in the limit $N \rightarrow \infty$: Almost every additive observable for which the local spin orientations are all identical features macroscopic fluctuations on a random symmetric state. The equality of local spin orientations is crucial here: If these orientations were chosen randomly and independently, the expectation value would hardly  fluctuate, since most Eigenvalues of the full variance-covariance-matrix $V$ Eq.~(\ref{VCM}) are typically small.

\section{Conclusions and outlook} \label{conclusions}
Many different approaches to entanglement eventually turn out to be equivalent, motivating the powerful concepts of entanglement monotone and entanglement measure \cite{Plenio}. 
Our results emphasize that macroscopic entanglement, as quantified by Eq.~(\ref{macrodef}), should never be treated as a synonym for a measure of entanglement: In particular, there are entangled states that feature vanishing macroscopic entanglement (Section \ref{geomwim}).  

Random physical states reflect this intricate relationship by their trajectory in the $(E_{\text{G}}, \mathcal{M})$-plane (Section \ref{randphysstse}), converging to Haar-random states, which feature large geometric and small macroscopic entanglement. The typical size of macroscopic superpositions of random physical states grows, but not as fast as the system size -- consequently, the normalized measure of macroscopicity converges to 0 in the limit $N \rightarrow \infty$. Symmetric states are naturally much less geometrically entangled and much more macroscopic, and it remains to be studied whether there are ensembles beyond symmetric states for which the actual spin orientations in the definition of the additive observables [Eq.~(\ref{addobs})] are irrelevant. Such ensembles would be experimentally  valuable due to their robustness.

Further quantitative insight in the relation between macroscopic and geometric entanglement is  desirable. A general bound on geometric entanglement as a function of macroscopicity (or vice versa) seems hard to obtain, since both quantities are defined via a maximization procedure.  We believe nevertheless that our bound in Section \ref{ffss} can be improved considerably, and that a proof for the extremality of $\ket{\Xi(\theta, \epsilon)}$ can be found. We did not find any relationship between the closest separable state $\ket{\phi_1, \phi_2, \dots , \phi_N}$ and the maximizing spin orientation $\{ \vec \alpha_1, \dots , \vec \alpha_N \}$; such deeper connection would be  valuable. 

Control schemes that optimize multipartite entanglement implicitly exploit the typicality of entangled states within the ensemble of pure states \cite{Platzer2010}. Our results suggest that control strategies that aim at a macroscopically entangled target state will not only be affected by decoherence, but the unitary evolution also needs to be tailored in a much more precise way: While the manifold of states that feature high geometric entanglement is very large, this is not true for macroscopic states. 

Finally coming back to our proposed analogy [Fig.~\ref{analogy}], our results suggest that macroscopically entangled states play the role of four-leaf clover: They do not appear spontaneously after some random process, but only as the result of some meticulously designed artificial evolution, such as in a quantum computer \cite{Shimizu2013}. Further investigations of other ensembles of pure quantum states, such as canonical thermal pure states \cite{Sugiura2012} and random matrix product states \cite{Garnerone2010a,Garnerone2010}, will eventually rigorously confirm or dismiss the analogy.

\vspace{.5cm}

\subsection*{Acknowledgements}
C.-Y.P., M.K., and H.J. were supported by the National Research Foundation of Korea (NRF) grant funded by the Korea Government (MSIP) (No. 2010-0018295). M.C.T and K.M. acknowledge financial support by the Danish Council for Independent Research and the Villum Foundation. M.C.T. was financially supported by the bilateral DAAD-NRF scientist exchange programme. C.-Y.P. thanks Jinhyoung Lee and Seokwon Yoo for granting access to the Alice cluster system at the Quantum Information Group, Hanyang University.  The authors thank Christian Kraglund Andersen, Ralf Blattmann, Eliska Greplova, Qing Xu and Jinglei Zhang for valuable comments on the manuscript.

%\bibliographystyle{h-physrev}
%\bibliography{RelevantLiterature}

\end{document}